\documentclass[runningheads]{llncs}

\usepackage{tikz}
\usepackage{amsmath}
\usepackage{graphicx}
\usepackage{caption}
\usepackage{xcolor}
\usepackage[T1]{fontenc}
\usepackage{xurl}
\usepackage[all=normal, tracking=tight]{savetrees}

\newcommand{\name}{LLMix\ }

\begin{document}

\date{}

\title{\name: Quantifying Mix Network Privacy\\ Erosion with Generative Models}
\titlerunning{Combining Leaked Information from Mixing Nodes}

\author{Vasilios Mavroudis\inst{1} \and Tariq Elahi\inst{2}}
\authorrunning{V. Mavroudis and T. Elahi}
\institute{The Alan Turing Institute \and
University of Edinburgh\\
\email{v.mavroudis@turing.ac.uk, t.elahi@ed.ac.uk}}
\maketitle
 
\begin{abstract}
Modern mix networks improve over Tor and provide stronger privacy guarantees by robustly obfuscating metadata. As long as a message is routed through at least one honest mixnode, the privacy of the users involved is safeguarded. However, the complexity of the mixing mechanisms makes it difficult to estimate the cumulative privacy erosion occurring over time. This work uses a generative model trained on mixnet traffic to estimate the loss of privacy when users communicate persistently over a period of time. We train our large-language model from scratch on our specialized network traffic ``language'' and then use it to measure the sender-message unlinkability in various settings (e.g. mixing strategies, security parameters, observation window). Our findings reveal notable differences in privacy levels among mix strategies, even when they have similar mean latencies. In comparison, we demonstrate the limitations of traditional privacy metrics, such as entropy and log-likelihood, in fully capturing an adversary's potential to synthesize information from multiple observations. Finally, we show that larger models exhibit greater sample efficiency and superior capabilities implying that further advancements in transformers will consequently enhance the accuracy of model-based privacy estimates.
\end{abstract}

\section{Introduction}
With the resurgence in popularity of Mix networks, due in part to the Snowden revelations, and upcoming real-world deployments like Nym~\cite{diaz2021nym}, HOPR~\cite{hopr2021}, and Elixxir it is increasingly critical and timely to have assurances that mix design privacy expectations stand up to reality. Mixnets protect against powerful adversaries that have complete surveillance of all traffic links, thereby enhancing the security assurances beyond those provided by Tor. Indeed, many traffic analysis attacks that work on Tor, e.g. website fingerprinting, are not readily applicable in mixnets as the timing and communication patterns are robustly obfuscated by the \emph{mixing strategies}. Furthermore, analytic solutions for larger mixnets are not tractable.
Thus most designers and operators carry out empirical privacy measurements, including  specific attacks that are critical to defend against, to determine the optimal mixing strategy and parameters for their use case~\cite{piotrowska2017loopix,diaz2021nym,diaz2004thresholdpoolmix,guirat2022mixnet}. The two most common metrics used are: entropy and likelihood difference \(\epsilon\). They are both used to statistically summarize the leakage of the network or a single node. However, by definition they do not combine observations from multiple mixing rounds and thus do not estimate the cumulative leakage over time (only the average), which is more realistic since a mixnet is expected to by used by a population over multiple rounds and a time horizon. Our work addresses this gap in the capabilities of existing tools, enabling operators to measure the resilience of their mixing strategies over several mixing rounds.

In particular, we introduce an automated design-agnostic approach for discovering \emph{information leakage} in mixing strategies without the need for concrete attack realizations. This approach complements existing tools, and shifts the focus from the ``prior'' knowledge of the adversary and the specifics of the system details towards the privacy goals of the system and the objectives of the adversary. Central to this approach are: 1) our methodology for generating machine learning-compatible privacy-measuring tasks that model the goals of the adversary, and 2) the use of a transformer model as empirical privacy estimators (with regards to the aforementioned tasks).

We first introduce an encoding that represents and captures all relevant mix node transmission metadata as a stream of messages and represents them in a format that is compatible with modern large language models. Using~\cite{kuhn2019privacy} we translate the high-level privacy goal, specifically sender-message unlinkability, into a privacy game and model it as two machine learning (ML) tasks (i.e. distinguishing between two senders, and one sender amongst many). We then generate network traces for a range of mixnet configurations and train the transformer model on them. During training the model learns the underlying rules of mixing by trying to generate valid mixnet traffic traces (i.e., guess the next message transmission). This traffic-aware model
is then used to solve our privacy-estimating tasks given a new mixnet traffic trace. 

Our model LLMix\footnote{We will open-source our trained generative models, tools and scripts upon publication.} is a transformer (two variants) trained \textit{from} \textit{scratch} to process and classify mixnet traffic. We evaluate various mixnet strategies and find that configurations that impose the same average latency are not always equally robust with regards to the unlinkability property (Section~\ref{subsec:exp1}). We focus on the leakage of individual nodes as our goal is to provide a best case (for the defender) privacy comparison of the different mixing strategies and parameters, without the added complexity of network topology choices. This setup is in keeping with the standard \emph{anytrust} assumption where even one mix node must be capable of ensuring the privacy of the users~\cite{wolinsky2012scalable,ma2022stopping}. We also train a larger variant of our model to study if it benefits from an increased observation window. We confirm our hypothesis and show there is a clear relationship between the number of messages captured and privacy loss (Section~\ref{subsec:exp2}). Comparing these results with classic statistical tools~\cite{piotrowska2017loopix,diaz2021nym,guirat2022mixnet} we show that our proposed technique provides a better privacy estimation (i.e. the privacy information leakage). Finally, provide initial evidence that that the number of \textit{learnable} parameters directly influences the model's sample efficiency (Section~\ref{subsec:exp3}).

Overall, this work realises a task-to-model framework enabling ML advances to be applied to harden mixing networks and guide parameterization. In particular, we:
\begin{itemize}
    \item Introduce a new traffic analysis task format that is solvable by language models and train from \textit{scratch} a mixnet traffic analysis tranformer model.
    \item Provide, to the best of our knowledge, the first design-agnostic generative model (LLMix) that accounts for cross-round leakage in mixing privacy estimations.
    \item Evaluate a range of configurations and find that some combinations achieve better privacy with the same latency.
    \item Study sample-efficiency, and show that larger models are likely to have better privacy-estimating capabilities.
\end{itemize}

\section{Background}\label{sec:background}

\subsection{Mix Networks}
First proposed as a mechanism for untraceable electronic mail~\cite{chaum1981mixnet}, Mix networks (or mixnets) have since been widely adopted for several applications including secure e-voting~\cite{kristian2012evote}, anonymous routing~\cite{chen2015mixrouting} and anonymous messaging~\cite{piotrowska2017loopix,alexopoulos2017mcmix}. HOPR~\cite{hopr2021} and Nym~\cite{diaz2021nym} are two recent examples of real-world deployments using the latest advances in mixnet techniques as stronger alternatives to more established solutions like Tor and VPNs.

To transmit a message over a Mixnet, the sender selects a route over a number of nodes (or hops) before the message arrives at the recipient. A single honest mix node in the route ensures the unlinkability of the message. Each message is padded to a constant length before it is sent and then cyrptographically processed at each hop (either decrypted or re-encrypted depending on the scheme used) to prevent traceable bit patterns, with an additional delay at each hop. The particular \textit{mixing strategy} dictates this delay and is what differentiates mixnets from Tor (uses FIFO routing). At the expense of additional latency, the delay ensures that multiple messages from (ideally) different users are co-resident in each node thus creating confusion for the adversary to trace the path of messages through the network, thus preventing the linking of sender and recipient. Dummy messages can further obfuscate the link between sender and recipient at the expense of bandwidth.

Mixing nodes are the building blocks of mixing networks as each of them individually ensures the realisation of the mixing strategy. In fact, a mix net with a single node would provide the best possible mixing of the traffic available. However, for scalability and to avoid single points of failure, most modern mixnet designs use a stratified topology where nodes are arranged in ordered layers with messages traversing the network across the layers in sequence. This fragments the traffic thus reducing the homogeneity of the mixing. Most of the mixnets using such a topology operate under the anytrust assumption\cite{wolinsky2012scalable} where at least one server in the user’s path must be honest. Thus security comes from distributing trust across many relay operators. %

\subsection{Transformer Models}\label{subse:llms}

Many different approaches to language modelling have been proposed~\cite{turian2010word,miklov2013words,pennington2014glove,peters2018elmo}, however massive scaling (i.e., large) in language model parameters has recently yielded unprecedented performance improvements across several tasks~\cite{devlin2018bert,brown2020fewshotgpt,rae2022scaling,yang2023harnessing,zhao2023survey}. 
Large language models (LLMs) typically use a \textit{transformer} neural network architecture~\cite{vaswani2017attention} designed to process sequential data such as natural language based on self-attention. Self-attention is an attention mechanism~\cite{bahdanau2015attention} that allows dependencies in sequential data to be modelled independently of their distance in the input and output. Importantly transformer models are highly parallelisable, enabling the scale necessary for LLMs. In what is termed ``autoregressive training'', LLMs for natural language tasks are usually trained to predict the next word in a given sentence. This process is performed iteratively, generating one word prediction in each step. Despite the relative simplicity of this process, autoregressive training is sufficient to capture much of the syntax and semantics of language. 
LLMs can generate coherent and contextually relevant natural language, allowing them to perform well in many previously unsolved tasks~\cite{dao2022flashattention,gpt4}. They are given a vocabulary set that defines all of the unique tokens the model can recognise and generate. Depending on how the vocabulary is defined, each token may correspond to a whole word, a subword, a character, or a byte. Tokenization splits raw text (e.g., a phrase, sentence, paragraph, or a document) into individual tokens from the vocabulary, used as input to the LLM. While transformers are generative models, they can nonetheless be used for classification tasks while also taking advantage of their ability to resolve long-range dependencies in sequential data. Section~\ref{sec:model} outlines how we train them from scratch to process non-human language i.e., mixnet network traces.

\section{Threat Model}\label{sec:threatmodel}

We assume the standard global passive adversary (GPA) that is able to observe all network traffic between users and mixnode under examination~\cite{piotrowska2017loopix,chaum1981mixnet,attarianMWB23,guirat2022mixnet}. The adversary observes a fixed number of network events i.e. messages entering and leaving the network. The messages are all indistinguishable from each other (Sphinx~\cite{danezis2009sphinx}). The adversary does not actively inject, drop, replay or delay messages, and does not operate any compromised end-users, i.e. sender or recipient of a message. The adversary has the ability to corrupt nodes, however, at least one of the mixnodes each message is traversing through is assumed to be honest. This is the anytrust assumption and is commonly used by many known designs (e.g. Vuvuzela~\cite{van2015vuvuzela}, Karaoke~\cite{lazar2018karaoke}, Loopix~\cite{piotrowska2017loopix}, Nym~\cite{diaz2021nym}). Corrupt nodes are assumed to operate on a FIFO manner as the adversary can fully observe their operation and hence they do not contribute to the mixing~\cite{piotrowska2017loopix}. 

\subsection{Adversarial Goals \& Tasks}\label{sec:tasks}
The overall goal of an adversary is to breach the privacy of the users, specifically to break the \textit{unlinkability} privacy property of the mixnode and link the communication between the sender and recipient of the message. We focus on this goal since it is fundamental to mix networks and the basis of other privacy goals. Kuhn et al.\ formalize this \textit{sender-message unlinkability} privacy property as the $SM\overline{L}$ notion, a fundamental desired property shared by many mixnet designs and thus of main concern when evaluating mixnet designs and configurations (e.g., Loopix~\cite{piotrowska2017loopix}). We choose to focus on this property due to its importance in most mixnet use cases. However, our methodology can be easily adapted to support any notions defined in~\cite{kuhn2019privacy}. 

Our work extends the foundational work of Kuhn et al. by translating their established privacy goal and concept framework into corresponding tasks within the realm of machine learning. We then use the success of the adversary at the ML task to gauge the privacy level of a given mixnet with respect to that goal. Mixnet designers and operators can use this information to then make design and deployment decisions such as rule out mixing strategies and configurations that provide subpar privacy. Note that in the literature the privacy ``game'' commonly involves guessing the sender of a \textit{single} message rather than identifying a \textit{persistent contact}. We consider the latter to be more pragmatic. In the former case, when the privacy leakage is minor, the measurement noise will not indicate that the adversary has gained any substantial advantage by observing a single message. However, leakage is accumulative and observing several messages may result in an advantage eventually, as our results show. We now introduce the task that instantiates $SM\overline{L}$:

\subsection*{Task: One of Two}\label{subsec:taskI}
Given a recipient $B$, the adversary aims to identify the user $A$ who sends messages to $B$.
The network has several actively communicating users but the adversary has to choose between \textit{two} potential senders (e.g., due to prior knowledge through external information). A coin-flipping adversary has a $\frac{1}{2}$ chance of guessing correctly in this task and the privacy loss is given by any adversarial performance that exceeds this mark.

\section{Converting Privacy Goals into ML Tasks}\label{sec:model}
As discussed in Section~\ref{sec:background}, transformers were designed to 
solve natural language processing and computer vision tasks. We argue
that network traffic exhibits properties and structure that is analogous to
those applications, thus making network traffic a suitable application area.
We now delve into the details of leveraging the capabilities of an LLM
to estimate the privacy of a given mixnet with regards to a specific adversary 
(i.e., desired privacy property and corresponding ML task).

\subsection{Network Traffic Encoding}\label{subsec:encoding}
\begin{figure}
\centering
\begin{minipage}{.5\textwidth}
  \centering
  \includegraphics[width=0.8\textwidth]{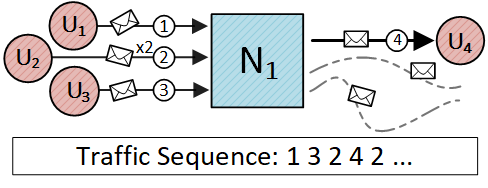}
\end{minipage}%
\begin{minipage}{.5\textwidth}
  \centering
  \includegraphics[width=1\textwidth]{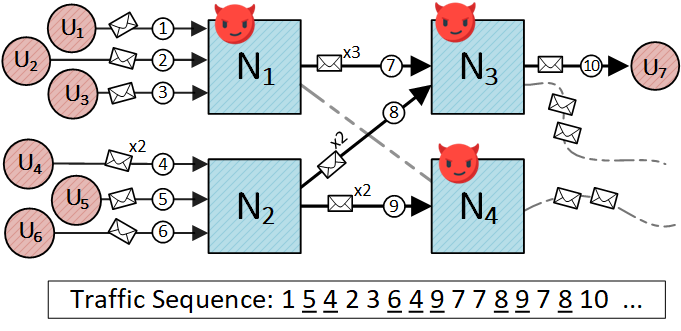}
\end{minipage}
\caption{Intercepted mixnet traffic can be efficiently represented as a sequence. Messages are replaced by the numerical id of the network link they traversed. The leftmost figure shows the traffic as it is routed through a single node, while the rightmost one shows a mixnet with several compromised nodes (anytrust assumption). The underlined link ids correspond to traffic that they adversary does not have full visibility on due to N$_2$ being honest.}
\label{fig:sequences}
\end{figure}

Modern mixing strategies and the use of cryptographic message formats (e.g., Sphinx~\cite{danezis2009sphinx}) leave only a few types of metadata exposed. Specifically, for each message transmission event, an observer knows either the sending or receiving systems/parties, the direction of the transmission, and the relative order of this transmission to the rest of the transmissions taking place in the network. Figure~\ref{fig:sequences} shows how an adversary can represent a snapshot of the network's activity as a sequence that encodes the aforementioned metadata.

In each case, the sequence produced retains all the information that a passive adversary can collect by eavesdropping the network links and is in a format compatible with LLM input sequences. This is because we have translated network traffic into a language with the following characteristics:
Each network event is represented by its corresponding link id.
Each ``word'' (i.e., link id) is unique and has a single meaning.
Two or more words can be combined to form a sentence (i.e., a traffic sequence) where the relative order of the events matters.

Given these characteristics and the ability of transformers to process natural language, we argue that
the network activity ``language'' outlined above is considerably easier for an LLM to learn (compared to a natural language).
Our encoding ensures there is no lexical ambiguity (i.e., one word with two meanings) and the vocabulary is relatively 
small for networks of moderate size. Transformers are mathematical functions and thus the input sequences must be converted 
into numbers. For natural languages, tokenizers are used to convert character-based words to integers. In our case, the mapping is even more straightforward as our link ids are already monotonically increasing integers (a word-level tokenizer). A ``0'' represents the absence of network activity.

Note that in NLP the encoding used to represent the information is always the same (e.g., voice or written text of a human language) regardless of the task at hand (e.g., classification, sentiment analysis, classification). This is an important characteristic of the way tasks are modelled to be solved by LLMs and part of why LLMs generalize so well. In the same vein, our network activity ``language'' is also generic as it represents the activity taking place in the mixing network in a lossless manner regardless of the end-goal of the adversary. This approach is unlike classic ML methodologies that derive features specific to the objectives of the analysis that the model then operates on (e.g., website fingerprinting~\cite{wang2014effective}).

\subsection{Privacy Properties \& Games}
In general, we fix the set of security or privacy properties a mix node will be evaluated against. As mentioned before, Kuhn et al.~\cite{kuhn2019privacy} provide a comprehensive framework that formalizes an anonymous network's privacy goals as \textit{notions} and defines a hierarchy between them. Given a particular use-case, the framework allows practitioners to define the communication setting and express their intuitive privacy goals as formal privacy notions and \textit{games}. Using this methodology, the steps of the game that corresponds our task above (Section~\ref{sec:tasks}) are: 1) the adversary picks two potential senders and a recipient  from the mixnet's population; 2) the challenger checks if the senders and the recipients are distinct; 3) based on a random bit \textit{b} the challenger inputs a scenario into the mixnet testbed; 4) the adversary observes the traffic sequence and outputs a `guess' \textit{b'} as to which scenario was executed.

\begin{figure}[h]
    \centering
    \includegraphics[width=0.5\textwidth]{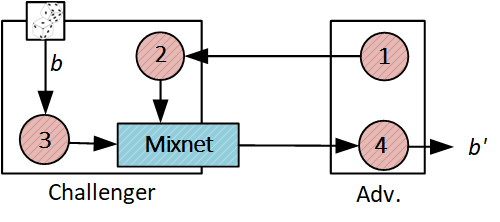}
    \caption{A sender indistinguishability game: 1) the adversary picks two potential senders and a recipient;
    2) the challenger checks the selections;  3) the challenger randomly chooses and runs the scenario; 
    4) the adversary observes the traffic sequence and outputs a `guess' \textit{b'}.}
    \label{fig:game}
\end{figure}

Once the game(s) has been defined, we use a simulator, or an emulator, of the network to generate
data from the scenario. From each run, we get the traffic sequence and ground-truth label (i.e., bit \textit{b} above).
The label is 0/1 depending on if user A or user B was the persistent contact. We can then compile a dataset (each row has a
traffic sequence and its label) and train models.

\section{Data Collection}\label{sec:setup}
Before we present our experimental results, we discuss our setup and the rationale for our design decisions for our testbed. 
The testbed realises mixing nodes and implements the threshold, pool, and Poisson mixing strategies. Our testbed is implemented 
in Python 3.12 and is comparable with those used in the mixnet literature (e.g. ~\cite{guirat2022mixnet,optimal-mix,piotrowska2017loopix}).

\subsection{Parameters}
The number of active users ($N$) is at least $3$ (i.e., two senders, one recipient) but in practice,
we study significantly greater user numbers. The rate of sending $r_i$ per user
is a ratio i.e., if $r_i=0.5$ then the user sends 0.5 messages in every unit of time.
The global sending rate $R\leq1$ (up to 1 message every time unit).
This means that time moves forward only when there is a message transmitted somewhere in the network.
In other words, the adversary will simply discard periods of inactivity from their dataset.
In cases where the users all behave uniformly, $R=r_i*N$. The average end-to-end latency 
is denoted $l$ and is not capped as it depends entirely on the mixing parameters
$\lambda$, $n$ and $p$ depending on the mixing strategy used. Note that $p$ is a ratio 
between the messages held in the pool and $n$. $p$ is always smaller than $1$
as the pool cannot be of equal (or larger) size to the threshold buffer.
Finally, the users need to have at least one contact otherwise the are not considered active and should not be accounted for in $N$.

\subsection{User Traffic}
Mixing networks rely on user traffic to provide anonymity. Consequently, 
a large number of users and hence more traffic makes it easier for defenders
to be assured of a high level of privacy~\cite{dingledine2006anonymity}. 

Threshold mixes have been typically studied on the assumption that all the users participate equally in each \textit{batch} of messages shuffled together.
As stated in the original paper introducing threshold mixes~\cite{chaum1981untraceable}:
``...each participant supplies the same number of messages to each batch''. 
Having perfectly balanced \textit{batches} where all the users are equally represented,
significantly strengthens the privacy of the network but the usability impact of such
a rule would make threshold mixes completely impractical to use. 
Instead, and more realistically, we assume that all active users send messages at approximately the same rate. 
This is favourable to the defender as it maximizes the sender diversity for the messages traversing the network 
and prevents users from trivially standing out. It also allows us to evaluate the quality of the mixing
while minimizing the noise that a skewed user traffic distribution would introduce.
In practice, real-life sending patterns are not perfectly uniform and thus the adversarial performance 
reported in Section \ref{sec:experiments} should be treated as an upper-bound for the privacy a 
given configuration can guarantee. Moreover, sophisticated (active) adversarial strategies
can exacerbate uneven sending distributions further so as to erode privacy even more.

We consider only two-party communication (cf. multicasting) via \textit{messages} at the application-layer.
We believe that this is representative of two-party exchanges that take place in short bouts of messaging activity.
The \textit{contact} of each user is chosen by sampling uniformly from the user list and does not assume any reciprocity 
(i.e., if A elects B as its contact, then B does not necessarily elect A too). This is to avoid introducing priors into the 
communication graph. %

Focusing on high-level messages, rather than network-layer packets, abstracts away from heavy hitters who send larger texts,
pictures, videos (that require multiple TCP/UDP packets) from standing out. In practice, cryptographic formats 
such as Sphinx~\cite{danezis2009sphinx} and techniques such as dummy packets prevent trivial deanonymization 
of heavy hitters, however some risk remains. %

Consistent with our threat model we do not consider active attacks such as
the $n-1$ message attack~\cite{serjantov2003activeatk}. Such attacks provide a significant advantage to the adversary but are more detectable and thus riskier, compared to our passive adversary.

\subsection{Burn-in Phase}
Before each testbed run, we perform a burn-in that aims to prefill the buffers and the pools of the mixing nodes. This ensures that the adversary gets to observe traces from a mixnet that is properly initialized. Poisson mixes are particularly sensitive to poorly initialized networks. To find the minimum time it takes for the buffers to stabilise, we tracked the number of messages in the mixnode buffers over time and for different $\lambda$
values. We found that after 300 timesteps all buffers are stably populated for $\lambda \leq 50$. We further corroborated the stability of the buffer and the mixing quality by measuring the buffer's entropy for over $1,000$ timesteps. Single-node threshold mixes without a pool do not require burn-in but we followed the same approach for threshold mixes with a pool.
Based on our findings, we set each burn-in to last 4096 timesteps followed by another period of random duration (between [1,4096]). This random component ensures that there is no alignment (e.g., traces always start after 96 messages have been transmitted from the network's exit node) between the traces generated. During this period, the ``suspect'' senders do not send any messages.

\section{Experimental Methodology}\label{sec:training}
For our experiments we use the Longformer~\cite{beltagy2020longformer} and train from scratch two variants (with moderate and large parameter number respectively). This is an important distinction as all pretrained large language models (LLMs) found online are trained on natural language tasks and do not transfer to network traffic.

Moreover, some of these model architectures (e.g., LLaMa) are very capable but also require an extremely high cost investment in compute to train them. We opted for the Longformer as it introduces a variation of self-attention (i.e., local attention) that scales linearly to the sequence length. It has approximately 149 million parameters and its primary advantage is that it supports considerably longer input sequences at a relatively low computational cost. In contrast, other self-attention mechanisms grow quadratically with the sequence length, making it infeasible to process long sequences. With local attention every token attends only to tokens in its vicinity defined by a window $w$. In particular, each token attends to the $\frac{1}{2}w$ preceding and the $\frac{1}{2}w$ succeeding tokens. Longformer combines this with \textit{dilation} allowing for increases in the window size without incurring additional memory costs. 

Despite its sophistication, the Longformer is a generative model and simply predicts the subsequent token based on the past. To use an LLM for classification, we follow the standard practice and replace the first token of every sequence with a special classification token ``[CLS]''. The final hidden state corresponding to this token is used as the aggregate sequence representation. This representation is is then passed through a simple classifier (usually a Linear layer) to get the predicted class label. This approach was first introduced in BERT~\cite{devlin2018bert} and is an established way of training an LLM for a classification task.

\section{Experiments}\label{sec:experiments}
In this Section, we use our model \name to evaluate the mixing strategies and parameters under various
circumstances and configurations. In all our experiments, we use our testbed to generate data from a 
network with a specified number of users, topology and mixing parameters. 
We conduct our experiments under the anytrust assumption which requires that as long as a message crosses one honest node its privacy should be preseved (see also Section \ref{sec:threatmodel}).
We split the produced labelled 
data into three datasets: \textit{training}, \textit{validation} and \textit{testing}.
All our reported results correspond to the models' performance on the testing dataset. Testing dataset 
samples were not encountered during training or when tuning our models' hyperparameters (validation dataset).

\subsection{Configuration \& Hyperparameters}\label{subsec:hyperparameter}
\name is a variant of the Longformer model with its parameters optimized for traffic 
analysis tasks and lower computational costs. We now describe the most important
parameter choices and our rationale. Note that many of the considerations and the practical
rules of thumb are specific to LLMs.

\noindent\textbf{Hardware.}
We use 1) 4 DGX-MAX-Q Nodes each with 8x Nvidia V100 GPUs with 32GB of VRAM,
and 2) an HGX100 GPU planar with 4x Nvidia A100 80GB GPUs.

\noindent\textbf{Software.}
We use Python 3.10.4 with the technically relevant packages being: 
PyTorch 2.0 backend, Transformers 4.30.2~\cite{wolf2019huggingface}, Accelerate 0.23.0\footnote{https://huggingface.co/docs/accelerate/index} and NumPy 1.26.0.

\noindent\textbf{Batch size.} The batch size governs the training speed of the model.
Very small values (e.g., 1,2) can result in extremely slow training that will not converge
within a reasonable timeframe. Thus larger values (e.g., 256, 512, 1024 or larger) are preferable
as they increase the rate of training and prevent instabilities during training~\cite{mccandlish2018empirical}.
Note that the relationship between the batch size and the training speed is not linear.
The maximum batch size supported by the V100 GPUs (i.e., GPU memory) was $32$. This value,
although moderate, is consistent with other works training Longformer models~\cite{beltagy2020longformer}.
 
\noindent\textbf{Gradient Accumulation.}
Gradient accumulation splits the batch of samples into mini-batches that are processed sequentially.
While the mini-batches are processed, the gradients of those steps are collected and accumulated without updating the model. The model is updated once all the mini-batches have returned. We use 8 gradient accumulation steps that bring the effective batch size to 256 (32x8). Understandably, this process introduces significant latency in the training process as it requires moving data in and out of the GPUs VRAM. However, it significantly improved the 
training stability of our models and minimized the instances of non-learning models when increasing the complexity
of the task in curriculum learning.

\noindent\textbf{Epochs.}
We did not define a set number of epochs but instead let the models train at each level of the curriculum
until there was no significant evaluation accuracy improvement (1\% or more) for at least 20 epochs.
This is standard practice and is implemented using an \textit{EarlyStoppingCallback} that takes as
parameters the number of epochs (20) and the minimum improvement difference (0.01).

\noindent\textbf{FP16 Training.}
We used half-precision floating-point numbers (i.e., FP16) as it allows for faster computations. 
The main speedup comes from storing the layer activations in 16-bit precision and thus making numerical operations easier. This also results in 16-bit gradients. However, this does not provide any memory usage benefits. This is because the model is saved in the GPU memory in \textit{both} 16-bit and 32-bit precision and can thus even use more memory that a non-mixed precision version. Moreover, the gradients are converted back to full precision for the optimization step thus preventing any potential memory savings. We experimented with various batch sizes and precision settings and in our setup FP16 allowed for faster computations without impacting the maximum batch size we could fit in the GPU's VRAM.

\noindent\textbf{Attention Window.} 
We set the attention window ($w$) to 128. The motivation for this was that we wanted each attention head to 
be able to observe a full message shuffling round if possible. In this case, the window allows for up to 
64 tokens observed before and 64 after the mixing. Thus a mixing round on a mix with threshold 50 will be
fully observable. Of course, LLMs employ multiple attention heads and combine observations from two or more
heads if needed. %
We validated this in our use case by setting the window to 64. The model was still able to 
analyse the traffic even when the threshold was larger then $\frac{1}{2}w=32$. Moreover, preliminary
experiments with threshold values that exceed the 128 token window also confirmed this.

\noindent\textbf{Vocabulary Size.}
This defines the maximum number of different tokens that can be represented as tokens in the LLMs' input.
We used a vocabulary size of 10,000 tokens as this was enough for the configurations we evaluated. In particular, the vocabulary size needed is about twice the number of network links.

\noindent\textbf{Model Architecture.}
Due to hardware constraints we also had to reduce the dimensionality of the 
encoding and the pooler layers to 128 (i.e., hidden size). A common value is 512 when processing 
longer natural language sequences. As well as, the number of hidden layers to 8.
Values between 8-12 are common for smaller models while values of up to 30 are
used for larger ones. Finally, we set the number of attention heads to 12
which is a commonly used value. We arrived at this combination of values through
a grid search of the parameter space, trying to balance learning performance and 
speed within the capabilities of our hardware.

\subsection{Latency vs. Privacy}\label{subsec:exp1}
We now focus on the three most common mixing strategies from the literature: 
Poisson mixing, Threshold mixing and Threshold mixing with a pool. These methods
all try to achieve the same goal (privacy) but their usability impact (latency)
is not identical. In this experiment, we study the level of privacy achieved by
a single node implementing one of the above mixing strategies with regards to the
average latency imposed to the messages traversing the network. Using a single node, 
allows us to isolate any impact network's topology might have to the results. 
The single node serves could also be considered as a high level abstraction of a 
whole mixnet where the adversary does not have access to intra-node communication links.

\begin{figure}
    \centering
    \includegraphics[width=0.95\textwidth]{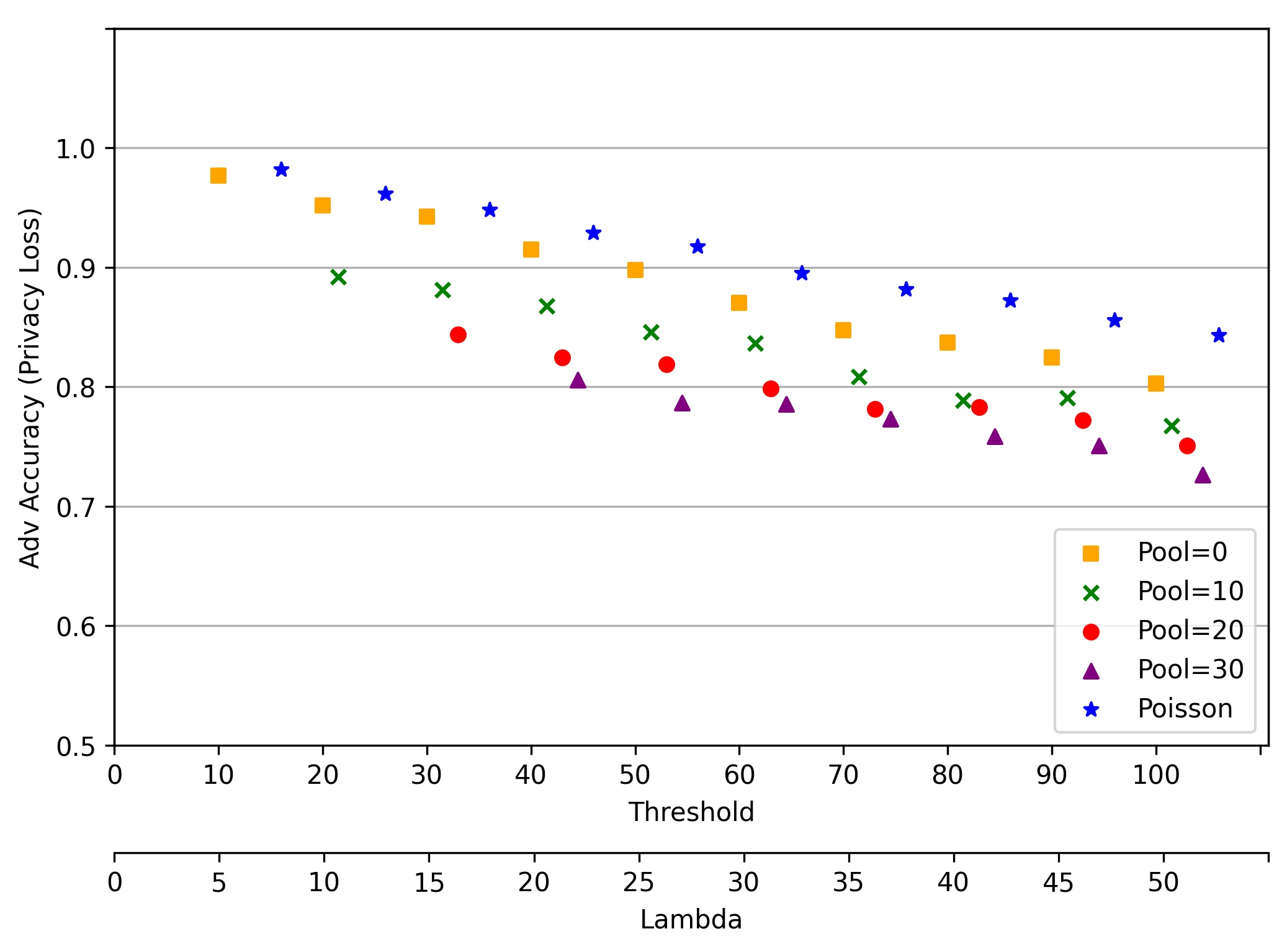}
    \caption{Privacy-loss estimation for different types of mixing strategies. The estimation task is guessing the sender of a message arriving to a given recipient. The total users are 100 and the possible senders are 2. The x axes are aligned by the latency each threshold/lambda value imposes. Lower values are better as values closer to 0.5 indicate better privacy. A node with a pool=0 is equivalent to a threshold-only node.}
    \label{fig:acc_1}
\end{figure}

We evaluate mixnodes with thresholds in the range of 10-100 and the corresponding (with regards to the latency 
imposed) $\lambda$ for the Poisson nodes in the range 5-50 (with a step of 10 and 5 respectively).
Our network has 100 users who each generate 1 message per 100 seconds. The privacy-estimation 
task assumes an adversary who given a recipient $r$ guesses $r$'s persistent contact between two 
possible senders (Section~\ref{subsec:taskI}).

We started by training one \name model on labelled data collected (2048 messages) from a mixnode
with threshold 10 and no pool. This initial training occupied a V100 GPU and took approximately 2
days before the LLM performed measurably better than a random guesser. We kept training the network
for another 2 days until it converged to its final performance as reported in Figure~\ref{fig:acc_1}.
Then we took advantage of the homogeneity of the datasets and used this trained model as 
a pre-trained basis for follow up training (curriculum learning). We trained another 43 models for 
for all the remaining mixing strategies and configurations (Figure~\ref{fig:acc_1}).

We observe that Poisson mixing performs worse than every other strategy and configuration.
This can be explained by the way \textit{continuous} mixing works. In particular, batch based
mixing strategies (e.g., threshold) do not flush messages out unless there are enough messages
in the pool. In contrast, continuous mixes favour usability (e.g., to prevent messages from
staying impractically long in the node) and treat every message independently. On average 
these two approaches ensure that each messages is shuffled with at $n-1$ other messages 
when in the node. However, there are edge cases where the mixing quality deviates from the 
average. As explained in Section~\ref{sec:threatmodel}, we assume a uniform sending behaviour 
for all the users which is a favourable assumptions for the defender. 

\begin{figure}
    \centering
    \includegraphics[width=0.95\textwidth]{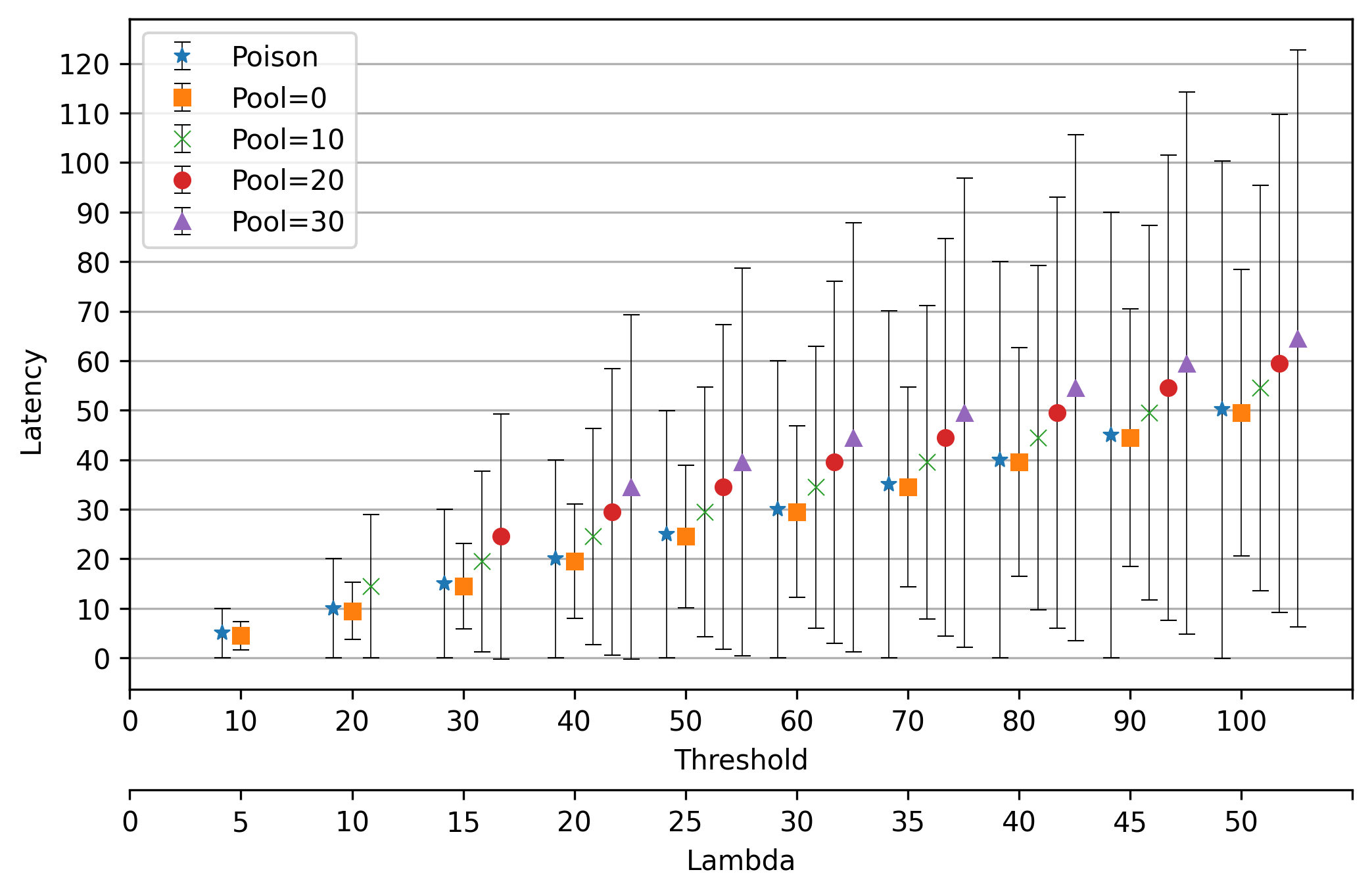}
    \caption{Mean latency and standard deviation (in seconds) for various configurations in Threshold, Pool and Poisson Mixes. We observe that 
    different configurations might have the same mean latency (variance might differ). A node with a pool=0 is equivalent to a threshold-only node.
    }
        \label{fig:latency}
\end{figure}

The number of messages in the buffer exhibits substantial variance which is the primary cause 
of the model's lower privacy estimate. Moreover, we also observe that a pool provides a significant
advantage to a threshold mix. To better evaluate the cost of a pool in the usability of the network,
we measured the latency of all the configurations in evaluation. Figure~\ref{fig:latency} shows that 
a mix node with a threshold $n=70$ and a pool 
holding 10 messages $p=14\%$ (setup A) provides about the same privacy with a threshold $n=100$ and 
no pool (setup B). Interestingly, setup A also imposes higher average latency to the messages
traversing the network while the increased latency variance introduced by setup A is less than that of B.
Furthermore, a node with $n=80$ and a pool holding 10 messages $p=13\%$ offers more privacy than 
setup A, lower average latency and comparable latency variance. Note that the latency in Poisson
nodes equals the $\lambda$ value, whereas the latency of threshold mixes is equal to approximately $n/2$.
This is because messages that arrive in the buffer later have to wait less time before the threshold is
flushed. In fact the first message will wait $n$ seconds (assuming an arrival rate of 1 second/message)
while the $n$'th message will not experience any delay. 

Overall, the trade-off curves (privacy vs. latency) for each mixing configuration
differ and thus a privacy evaluation is useful to identify the setups that strike the 
best balance. Moreover, none of these configurations provides perfect privacy under the anytrust assumption (see Section~\ref{sec:threatmodel}).
Note that, if we assume a weaker adversary that does not have access to intra-node network links, a mixing network can be abstracted as a single mixing node as long as the correspondence between the single node and the multi-node setup mixing is accurate.

\subsection{Num of Observations \& Model Size}\label{subsec:exp3}
One of our hypotheses is that advancements in the capabilities of transformer models will 
translate directly into better privacy estimations. To test this hypothesis we now train and evaluate a larger model. We use the same setup with Experiment 1 (Section~\ref{subsec:exp1}). To train the larger model we used Nvidia A100 80GB GPUs as the increased number of parameters did not fit in the original V100 with 32GBs of VRAM. However, as the models grow small batch sizes do not suffice for the model to learn. We thus increased the batch size to 200 but disabled the gradient accumulation as we found that it slowed training substantially. The increased attention window allows one head to observe more network events which together with the larger number of observations (4096 vs 2048) could give to the model greater inference capacity. Note that by increasing the observation length of each sample, we double also the space it occupies in the GPU's memory.

Table~\ref{tbl:exp5b} shows the accuracy of this model for sequences with various observation lengths. However, trained models require a fixed input length and will not work if samples of a different size are provided. As our larger model was trained on samples of 4096 events, we have to provide compatible samples. To achieve this, we used the same testing dataset (never seen during training). For each sample in this dataset, we randomly selected a region with the size we wanted to test against, and masked the rest of the events with zeros (`0' represents no network activity). 

From Table~\ref{tbl:exp5b}, we observe that indeed the larger model is more capable than the moderate sized-one. In particular, for a sequence of length 2048 the larger model achieves an accuracy of 0.88 while the smaller model reaches an accuracy of about 0.81~\ref{fig:acc_1}. This shows that not all the performance gain is due to the increased observation length. However, longer sequences indeed allow the model to do even better (0.95). In fact, the accuracy of the model depends heavily on the number of messages sent by the targeted sender (or observation length), and the performance declines as the sequences half in size. Interestingly, even 1 message sent allows the model to guess correctly with odds considerably better than random (accuracy 0.58).

\begin{table}[h]
\centering

\addtolength{\tabcolsep}{4pt}    
\begin{tabular*}{\columnwidth}{cc|ccc}

\multicolumn{2}{c}{\textbf{Sequences}}& \multicolumn{3}{c}{\textbf{Metrics}} \\
\# Obs & \# Messages & Adv. Advantage (ours) & Likelihood Diff. & Entropy\\
\hline \\
4096 & 20.7 & 0.958$^*$ & 0.269 & 5.824$^*$\\
2048 & 10.9 & 0.884$^*$ & 0.270 & 5.833$^*$\\
1024 & 5.1  & 0.776$^*$ & 0.270 & 5.849$^*$\\
512  & 2.0  & 0.661$^*$ & 0.260 & 5.865$^*$\\ 
256  & 1.0  & 0.583$^*$ & 0.262 & 5.859\ \ \\\\
\end{tabular*}

\caption{The relationship between the observations of the adversary, the messages sent by the real sender to the target recipient, the estimates of our metric, as well as the entropy and the likelihood difference \(\epsilon\) measured. The network has 100 active users and a threshold of 100. Adversarial advantage values closer to 0.5 indicates better privacy. We use * to mark the mean values that are statistically significantly (p-value 0.05) different from their adjacent means (rows above and below).}
\label{tbl:exp5b}
\end{table}

\subsection{Comparison with other metrics}\label{subsec:exp2}
Entropy and likelihood difference \(\epsilon\) have emerged as standard metrics to quantify the adversary's 
advantage~\cite{serjantov2003mixentropy,diaz2004mixentropy,guirat2022mixnet,piotrowska2017loopix,piotrowska2021measurenym}. 
We now show that these measures have limitations and discuss how our proposed model provides more accurate estimations.

The entropy metric quantifies the information contained in the anonymity probability distribution of possible senders and receivers, of a given message, 
as viewed by an adversary. The effective anonymity set size is defined as the entropy of the anonymity probability distribution, which 
can be interpreted as the number of additional bits needed to uniquely identify the specific sender or receiver of a chosen message. 
Compared with the anonymity set size, entropy provides a probabilistic measure of the information that an adversary can determine from a sequence. 
More formally, let $X$ be a discrete random variable over the finite set X with probability mass function p(x)=Pr(X=x). The Shannon entropy
H(X) of a discrete random variable X is defined as:
\begin{equation}
    H(X)=-\sum_{x\in X}p(x)logp(x)
\label{eq:entropy}
\end{equation}

The likelihood difference \(\epsilon\)~\cite{piotrowska2017loopix} provides the expected difference in likelihood that a message leaving a specific terminal mix node is sent from one sender in comparison to another. In comparison to entropy, likelihood difference \(\epsilon\) focuses on the ratio between the probabilities that a selected message was sent by one of two chosen senders. It thus allows anonymity to be quantified more pessimistically for a strong adversary who is assumed to know a priori that one of two specific senders are communicating with a certain recipient. Given a message, its likelihood difference is calculated as: $\epsilon=|log(p_0/p_1)|$ , where $p_0=Pr[U_0]$ and $p_1=Pr[U_1]$ denote the probabilities of users $U_0$ and $U_1$ being the senders of that message respectively.

Continuing with Table~\ref{tbl:exp5b}, we examine the likelihood difference between sequences with a varying number of messages from the real sender to the target recipient. The mean values shown in the table correspond to 5 rounds (with 256, 512, 1024, 2048, 4096 observations) of 1,000 sequences each.
Each metric should capture the additional privacy loss as the number of messages increases. However as seen in the table, the difference between the 
256-512, 512-1024, 1024-2048, 2048-4096 is not statistically significant and thus the likelihood difference \(\epsilon\) fails to capture the extend of the leakage. Moreover, the difference between 256-2048 was not statistically significant either, while the reported values are very close to what~\cite{piotrowska2017loopix} considers safe. In comparison, our model shows that there is a clear leakage in each of these cases and
all mean differences were statistically significant.

The entropy metric captures some of the leakage but does not reveal its full extend. To understand why, we use an example from
\cite{guirat2022mixnet}. An entropy of 10 bits for a mixed message arriving at a recipient, indicates that the message is ``as anonymous as if it
was perfectly indistinguishable among about a thousand ($2^{10}$ = 1024) other messages''. If the same user sends a follow 
up message that mixes in the exact same way with messages from the same recipients, there is no additional privacy loss.
However, in practice the subsequent message will be shuffled with a different mixture of messages from different senders. 
Even if the entropy of the subsequent message is also 10 bits, the probability distributions of the two messages will differ e.g., allowing the adversary rule out combinations of senders and recipients (something our model is capable of). Subsequent messages will provide additional information that will allow further inferences. In our case all the entropy measurements were between 5.824 (for 4096) 5.859 (for 256) which corresponds to the messages being indistinguishable between 56.64 and 58.04 other messages respectively (at most 1/56.64 changes of guessing correctly the sender of a message). Entropy thus captures some of the leakage but as our model shows it under-reports its full extent. Interestingly, all the mean differences (except for 256-512) were statistically significant. The reason for this that the are only 1 and 2 leaky messages respectively and consequently only 1-2 low entropy measurements that did not provide enough signal. 

Overall, we conclude that while statistical tools are easy to use for debugging and quickly iterating when parameterizing a node, they fail to capture the full extent of the leakage. This gap in capabilities can be addressed by our proposed methodology and model.

\section{Related Work}\label{related-work}

The problem of traffic analysis for privacy purposes remains open and is a relatively popular theme in anonymous communication networks research. However, there are only a few sub-areas where ML has been actively used so far. 

Website fingerprinting attacks against the Tor have employed standard machine learning techniques for classification such as k-NN~\cite{wang2014effective}, Support Vector Machines~\cite{panchenko2016website}, random forests~\cite{hayes2016k}, and more recently convolutional deep neural networks~\cite{sirinam2018deep,bhat2019var}. In webpage fingerprinting, Danezis et al.~\cite{danezis2009traffic} used a Hidden Markov Model to fingerprint pageloads from a static dataset. Miller et al.~\cite{miller2014know} fit a Gaussian distribution on their dataset while Dubin et al.~\cite{dubin2017know} used deep learning to fingerprint traces from video streaming services.

Attack-agnostic traffic analysis for mixnets using modern ML models remains an underexplored area as tools and techniques from other types of anonymity networks are not applicable. Past works have focused on studying specific attacks against specific mixing strategies~\cite{statistical-disclosure,least-squares,statistical-disclosure-overview}, covering the range of different mix types from threshold~\cite{least-squares}, to pool~\cite{least-squares}, to continuous~\cite{danezis2004traffic}, and consider enhancements such as dummy traffic as well~\cite{dummies}. Other works aim to provide optimal parameters to secure continuous mixnets~\cite{guirat2022mixnet,optimal-mix} like Loopix. This work can be viewed as an extension to these tools and can be used in conjunction with them. This has the potential to more fully cover relevant sources of information leakage that the more limited measurements in the related work might miss.

\section{Conclusion}
We introduced a framework that allows system designers and operators to 
estimate the privacy of their mixing strategies against practical adversaries. Our approach defines a ``language'' and a task format that are compatible with modern large language models and thus makes it straightforward for all advances in language models to be directly incorporated in privacy estimation. We believe that such estimations are close to the actual advantage of a sophisticated adversary, and as models advance, we expect the gap to become even narrower. While the list of adversarial tasks can not be exhaustive, privacy goal systematization efforts (e.g.~\cite{kuhn2019privacy}) can guide operators towards identifying and ``taskifying'' their primary privacy objectives. With this framework in place, it is now straightforward for designers and operators to evaluate their configurations against the latest language models, under conditions and assumptions that fit their use cases.

\bibliographystyle{plain}
\bibliography{refs}

\end{document}